\documentclass[aps,prl,twocolumn,groupedaddress,showpacs]{revtex4}
\usepackage{graphicx}% Include figure files
\usepackage{dcolumn}% Align table columns on decimal point
\usepackage{bm}% bold math

\begin{document}

\title{Can we trust semiclassical description of particle creation?}

\author{Hrvoje Nikoli\'c}
\affiliation{Theoretical Physics Division, Rudjer Bo\v{s}kovi\'{c}
Institute,
P.O.B. 180, HR-10002 Zagreb, Croatia.}
\email{hrvoje@thphys.irb.hr}

\date{\today}

\begin{abstract}
The predictions of the semiclassical description of particle creation 
based on QFT in classical backgrounds may be significantly modified
when the source of the classical background is also quantized
and backreaction is taken into account. In the cases of a stable 
charged particle, expanding empty (Milne) universe, and de Sitter
universe with a true cosmological constant, the semiclassical particle
creation is completely blocked up.
\end{abstract}

%Semiclassical theories and applications, Quantum field theory in curved spacetime

\maketitle

Quantum field theory (QFT) is an appropriate theoretical framework for describing 
particle creation and destruction. In such processes energy must be conserved,
implying that the creation of a new particle is allways accompanied by the 
destruction of an old particle or a transition of the old particle to a lower state.
Such processes are most successfully described in perturbative QFT in terms of Feynman diagrams. Nevertheless, in some cases the perturbative methods are not very efficient, forcing us to use different types of approximations.  

One such approximation widely used for description of particle
creation is the {\em semiclassical} approximation. In such an
approximation, only the new created particles are described by QFT, 
while the role of the ``old" particles is approximately
described by a {\em classical} source field. 
The best known examples of such a semiclassical
description of particle creation are the Schwinger effect
\cite{schw,man,brout} in which a classical static electric field
causes production of electron-positron pairs, the 
Hawking effect \cite{hawk,bd,brout} in which the classical gravitational 
field of a black hole causes production of particles with a thermal 
distribution of energies, and particle creation caused by a time-dependent
classical gravitational field induced by the universe expansion \cite{park,bd}.
The best known specific example of the latter is particle creation
from a horizon of de Sitter universe \cite{padm}, which contains only a positive
cosmological constant, but not ordinary matter. The Schwinger and Hawking effect 
are actually closely related \cite{brout}. An even closer relation 
exists between Hawking effect and particle creation by de Sitter universe,
as in both cases it is the horizon that is responsible for 
particle creation with a thermal distribution \cite{padm2}. 
Another related effect is the Unruh effect \cite{unruh,bd}, according to which
a particle detector accelerated in vacuum detects particles.

All these effects are usually treated in a {\em fixed} classical background,
thus violating conservation of energy. To fix this problem, 
one has to study the backreaction of created particles on the source.
If the source (i.e., the classical background) is still treated classically,
one usually finds that the backreaction does not significantly 
influence the original particle creation with a fixed background.
Nevertheless, the real source is allways a quantum object, so, in general,
we cannot be sure that the semiclassical approximation is satisfying.

As a simple example, consider the Schwinger effect in which
the source of the electric field is a stable particle, say a proton.
According to the semiclassical analysis, the static electric field
should create electron-positron pairs. (In the case of a proton
the electric field is too weak to provide a significant rate of production,
but for the sake of argument one may consider a hypothetic particle 
having the same mass as proton, but a much larger charge producing a much stronger
electric field.) Energy conservation implies that, after the pair creation,
the energy of the proton and its electric field must be smaller than
that in the initial state, which is impossible since the proton
is a stable particle. Consequently, in such a case there can be no pair creation,
so the semiclassical analysis fails completely.

The purpose of this paper is to study in more detail how, in general, 
the quantum treatment of the source modifies the semiclassical 
treatment of particle creation. Our analysis represents 
a further development and generalization of a recent analysis \cite{nikblh} introduced to study Hawking radiation. 

The most convenient method for a semiclassical description of
particle creation by a classical source is the Bogoliubov transformation.
In general, one finds that the initial vacuum $|0\rangle$
transforms to
\begin{equation}\label{semicl}
|0\rangle \rightarrow \sum_e\sum_i c_{e,i} |e,i\rangle ,
\end{equation}
where $|e,i\rangle$ are energy eigenstates with the energies $e\geq 0$,
the label $i$ labels different states having the same energy, and
$c_{e,i}$ are the semiclassical probability amplitudes 
for particle creation, satisfying the normalization condition
\begin{equation}\label{norm}
\sum_e\sum_{i} |c_{e,i}|^2 = 1 .
\end{equation}
The state $|0,i\rangle \equiv |0\rangle$ is the unique vacuum, 
while states $|e,i\rangle$ with $e>0$ contain one or more particles. 

Clearly, energy is not conserved in (\ref{semicl}).
To fix that problem, we introduce the energy eigenstates 
$|E,I\!>$ of the source. Now in this fully quantized 
description of particle creation, the initial state
is not the vacuum, but a linear combination of these
source energy eigenstates. For simplicity, we assume
that the initial state is some energy eigenstate $|E_0,I_0\!>$.
Then the particle creation can be described as a transition 
\begin{equation}\label{q1}
|E_0,I_0\!> \otimes |0\rangle \rightarrow
\sum_E\sum_I\sum_e\sum_i D_{E,I;e,i}^{(E_0,I_0)}  
|E,I\!> \otimes |e,i\rangle ,
\end{equation}
where $I=1,\ldots,N_E$, and $N_E$ is the number of states 
with energy $E$.
The transition of a more general initial state can be easily obtained
from (\ref{q1}) by linearity of quantum mechanics.
Energy conservation implies that the amplitude $D_{E,I;e,i}^{(E_0,I_0)}$
must have the form
\begin{equation}
D_{E,I;e,i}^{(E_0,I_0)} = \delta_{E_0,E+e} d_{e,I,i} , 
\end{equation}
where, for the sake on notational simplicity, the dependence of $d_{e,I,i}$
on $E_0,I_0$ is suppressed. Therefore, (\ref{q1}) becomes
\begin{equation}\label{q2}
|E_0,I_0\!> \otimes |0\rangle \rightarrow
\sum_I\sum_e\sum_i d_{e,I,i} 
|E_0-e,I\!> \otimes |e,i\rangle .
\end{equation}

Strictly speaking, we cannot conclude anything more about the amplitude
$d_{e,I,i}$ without knowing the details of the quantum theory
of the source. 
Nevertheless, the knowledge of the semiclassical approximation
suggests that (up to a phase) $d_{e,I,i}$ should be approximately
proportional to the semiclassical amplitude $c_{e,i}$. Namely,
it is reasonable to use the approximation
\begin{equation}\label{e1}
d_{e,I,i}=\frac{e^{i\varphi_{e,I}} c_{e,i}}{\sqrt{{\cal N}}} ,
\end{equation}
where $\varphi_{e,I}$ are some phases. The factor $1/\sqrt{{\cal N}}$
is determined by the normalization condition
$\sum_e\sum_I\sum_i |d_{e,I,i}|^2 =1$. This leads to
\begin{eqnarray}\label{e2}
{\cal N} & = & \sum_e\sum_I\sum_i |c_{e,i}|^2 =
\sum_e N_{E_0-e} \sum_i |c_{e,i}|^2 \nonumber \\
& = & \sum_e N_{E_0-e} P_e = \langle N \rangle , 
\end{eqnarray}
where 
\begin{equation}
P_e=\sum_i |c_{e,i}|^2 
\end{equation}
is the semiclassical probability that the energy of created particles 
is equal to $e$. The quantity
$\langle N \rangle$ is the average number of source states having the
same energy, with the average being defined with respect to the semiclassical
probability $P_e$. 
Defining
\begin{equation}
|E_0-e\!> \equiv \sum_{I=1}^{N_{E_0-e}} \frac{e^{i\varphi_{e,I}}}{\sqrt{N_{E_0-e}}}
|E_0-e,I\!> ,
\end{equation}
(\ref{q2}) can be written as
\begin{equation}\label{q3}
|E_0,I_0\!> \otimes |0\rangle \rightarrow
\sum_e |E_0-e\!> \otimes \sum_i \sqrt{\frac{N_{E_0-e}}{{\cal N}}}
\, c_{e,i} |e,i\rangle .
\end{equation}

Now, to see the relation with the semiclassical result (\ref{semicl}),
consider the case in which $N_E$ does not depend on $E$, i.e.,
$N_E=\langle N \rangle$. Now (\ref{q3}) reduces to
\begin{equation}\label{q3.1}
|E_0,I_0\!> \otimes |0\rangle \rightarrow
\sum_e |E_0-e\!> \otimes \sum_i c_{e,i} |e,i\rangle .
\end{equation}
Thus, the probability that the created particles will be found in the state
$|e,i\rangle$ is equal to $|c_{e,i}|^2$,
which is the same result as the one
obtained from the semiclassical description (\ref{semicl}).
Furthermore, if energy $e$ is measured, then (\ref{q3.1}) collapses to
\begin{equation}\label{c1}
|E_0-e\!> \otimes \frac{1}{\sqrt{P_e}} \sum_i c_{e,i} |e,i\rangle ,
\end{equation}
where $1/\sqrt{P_e}$ is the appropriate normalization factor.
Similarly, the semiclassical theory based on (\ref{semicl}) implies
the collapse to
\begin{equation}\label{c2}
\frac{1}{\sqrt{P_e}} \sum_i c_{e,i} |e,i\rangle ,
\end{equation} 
which is nothing but the particle-creation part of the fully-quantized wave function
(\ref{c1}). This shows that the case $N_E=N$ restores the semiclassical
result. From this we conclude that {\em the semiclassical approximation
can be trusted when the number of the relevant source states with energy 
$E$ does not significantly depend on $E$.} 
This condition is expected to be satisfied
when the source is in a highly excited state, with the energy $E_0$ 
much larger than typical energies $e$ of created particles. 
Indeed, such a highly excited state justifies the treatment of the source
as a classical object. 

As an extreme deviation from the condition above, consider the case 
$N_{E_0-e}=0$ for $e>0$. Then (\ref{q3}) implies that the 
transition amplitude vanishes.
In this case the source with the initial energy $E_0$ cannot
jump to a lower state simply because such a state does not exist.
Consequently, a particle with energy $e$ cannot be created, despite
the fact that the semiclassical amplitude $c_{e,i}$ does not vanish.
Indeed, this is exactly why the electric field of a stable charged particle discussed in the introduction cannot create electron-positron pairs,
despite the fact that it should create them according to the
semiclassical Schwinger effect.

Now we see that the semiclassical approximation (\ref{semicl})
cannot allways be trusted. However, the essential effects of quantization
of the source can be caught without knowing all the details of the fully
quantized theory. Namely, (\ref{q3}) shows that (\ref{semicl}) 
can be improved by replacing the semiclassical transition
probability $|c_{e,i}|^2$
by the improved semiclassical transition probability 
$(N_{E_0-e}/{\cal N}) |c_{e,i}|^2$. 
Hence, we can replace (\ref{semicl}) by the
{\em improved semiclassical approximation}
\begin{equation}\label{semiclim}
|0\rangle \rightarrow \sum_e\sum_i 
\sqrt{\frac{N_{E_0-e}}{{\cal N}}} 
\, c_{e,i} |e,i\rangle .
\end{equation}
We see that
it is a very useful approximation, because it requires a minimal
knowledge on the quantum structure of the source.
All we have to know is the number $N_E$ of states with energy $E$.
We have written all the expressions for the discrete spectrum of energies,
but it is obvious how to modify them for the (perhaps more interesting)
case of continuous energy spectrum.

Now let us discuss various examples. 
We start with the Schwinger effect. This effect has been confirmed
experimentally by experiments involving electric fields produced
by heavy ions \cite{wright}. 
A nucleus of a heavy ion, of course, can decay
into a large number of lower energy states, which explains why the semiclassical
approximation is justified in this case. If, however, the source of the 
same electric field was a stable elementary particle
(the Standard Model of elementary particles does not contain such a particle,
but this is not relevant for our theoretical argument), our analysis shows
that in this case the Schwinger effect would be completely blocked up.

Perhaps the most interesting example is Hawking radiation
from a black hole (BH). This case is studied in more detail in
\cite{nikblh}. It is shown how the fact that the number
of quantum BH states decreases as the BH mass $M$ decreases (due to 
the evaporation) resolves the famous BH information puzzle,
making complete evaporation consistent with unitarity.
We also note that, although the semiclassical Hawking temperature $T=1/8\pi M$ is infinite for $M=0$, Hawking radiation stops when the BH mass drops to $M=0$,
simply because lower-energy BH states do not exist. 

Next we consider the Unruh effect. The energy source is provided
by the external force that accelerates the detector \cite{moch},
which effectively means that the source can be viewed as always
being in the same state. Consequently, our analysis does not modify
the semiclassical result. Nevertheless, we stress that physical
details specifying a realistic quantum detector may significantly
modify the semiclassical result \cite{marz}.

Now let us consider particle creation by the universe expansion.
At the semiclassical level, the metric must satisfy the 
semiclassical Einstein equation
\begin{equation}\label{einst}
G_{\mu\nu}=8\pi [ T_{\mu\nu}^{\rm bulk} + 
\langle\psi|\hat{T}_{\mu\nu}^{\rm creat}|\psi\rangle ] ,
\end{equation}
where $G_{\mu\nu}$ is the classical Einstein tensor, 
$T_{\mu\nu}^{\rm bulk}$ is the energy-momentum of the bulk matter described classically, 
$\langle\psi|\hat{T}_{\mu\nu}^{\rm creat}|\psi\rangle$ is the average
energy-momentum of created particles described quantum mechanically,
and the Newton gravitational constant is set to 1.
If the particle creation takes place, then the last term is not conserved,
i.e.,
\begin{equation}
\nabla^{\mu} \langle\psi|\hat{T}_{\mu\nu}^{\rm creat}|\psi\rangle \neq 0.
\end{equation}
Then the Bianchy identity $\nabla^{\mu} G_{\mu\nu}=0$ and the
Einstein equation (\ref{einst}) imply
\begin{equation}\label{cons}
\nabla^{\mu} T_{\mu\nu}^{\rm bulk} \neq 0.
\end{equation} 
In other words, the only possible source of energy for particle creation
is the bulk matter. Indeed, this is analogous to the energy source
responsible for Hawking radiation. However, if the initial universe
does not contain bulk matter, i.e., if its initial state is the vacuum
state, then a bulk-matter state with an even lower energy does not exist.
Consequently, particles cannot be created in an empty
universe without initial matter. On the other hand, the semiclassical
analysis leads to the result that a time-dependent empty (Milne) universe does lead to particle creation \cite{bd}. Therefore, the Milne universe
is another example in which the semiclassical particle creation
is completely blocked up in a fully quantum treatment.

Finally, let us consider particle creation by de Sitter universe.
At the semiclassical level this effect is very similar to 
Hawking radiation because in both cases it is the existence
of the horizon that is responsible for it. Nevertheless,
the energy source is very different from that in the case
of Hawking radiation. In the de Sitter case, the source is described
by a bulk energy-momentum that has a cosmological-constant form
\begin{equation}
T_{\mu\nu}^{\rm bulk} =\lambda g_{\mu\nu}.
\end{equation} 
Consequently, (\ref{cons}) implies
\begin{equation}
\partial_{\nu} \lambda \neq 0.
\end{equation}
In other words, particle creation by de Sitter universe is only possible
if the cosmological ``constant" is not a true constant, but can decay. Thus, 
particle creation by de Sitter universe with a true cosmological
constant is our final example of semiclassical particle creation completely 
blocked up in a fully quantum treatment. Nevertheless, it does not mean
that the universe with a realistic cosmological ``constant"
does not produce particles, because there are many {\em dynamical} models
of dark energy in which $\lambda$ is a dynamical quantity.
In fact, a recent result based on linearized quantum gravity \cite{nikqgcc}
indicates that a non-dynamical bare cosmological constant does not contribute
to the observed cosmological ``constant" at all.

To conclude, we have seen that the semiclassical description of 
particle creation can be trusted only if the number of source states
with given energy does not significantly depend on energy.
Otherwise, the semiclassical approximation should be replaced
by an improved semiclassical approximation (\ref{semiclim}).
When the source does not contain a state with an energy lower
than the initial one, then the semiclassical particle
creation is completely blocked up. The examples are
a stable charged particle that blocks up the Schwinger effect, 
the expanding empty (Milne) universe that blocks up particle creation
caused by the time dependence of the metric, and de Sitter
universe with a true cosmological constant that blocks up
particle creation caused by the existence of a horizon. 

This work was supported by the Ministry of Science of the
Republic of Croatia under Contract No.~098-0982930-2864.

\end{document}